\documentclass[preprint,tightenlines,eqsecnum,floats,aps,amsmath,amssymb,nofootinbib,prd,showpacs]{revtex4}

\usepackage{amssymb}
\usepackage{stmaryrd}
\usepackage{amsmath}
\usepackage{amsfonts}
\usepackage{mathrsfs}
\usepackage{CJK}
\usepackage{amsmath,amssymb,amsfonts}
\usepackage{graphicx}
\usepackage{subfigure}
\usepackage{color} 

\def\be{\begin{equation}}
	\def\ee{\end{equation}}
\def\ba{\begin{eqnarray}}
	\def\ea{\end{eqnarray}}
\def\nn{\nonumber}

\newcommand{\eqnref}[1]{(\ref{#1})}

\begin{document}
	
	\title{Topological classes of black holes in de-Sitter spacetime}
	
	\author{Yongbin Du}
	\affiliation{Department of Physics, South China University of Technology, Guangzhou 510641, China}

	\author{ Xiangdong Zhang\footnote{Corresponding author. scxdzhang@scut.edu.cn}}
	\affiliation{Department of Physics, South China University of Technology, Guangzhou 510641, China}

	\date{\today}
	
	
	\begin{abstract}
		
		In this paper, we investigate the topological number of de-Sitter black hole solutions with different charges $(q)$ and rotational $(a)$ parameters. By using generalized free energy and Duan's $\phi$-mapping topological current theory, we find that the topological numbers of black holes can still be classified as three types. In addition, we interestingly found the topological classes for de-Sitter $($dS$)$ spacetime with distinct horizon, i.e, black hole event horizon and cosmological horizon, will be different. Moreover, we also investigate topological classifications of dS black hole solutions in higher dimensions with or without Gauss-Bonnet term.
	\end{abstract}
	\maketitle

	\section{INTRODUCTION}
	
	Black hole is one of the most fascinating objects
	in universe and has receiving increasingly attention both in theoretical and observational physics. The classical properties of black holes can be well described by general relativity (GR). However, singularity theorem \cite{Penrose} indicates that GR has intrinsic limitation. Particularly, when one taking quantum effect into consideration, people found that black hole is not black at all \cite{Hawking} and surprisingly radiate like a black body and have rich thermodynamical properties \cite{thermo1,thermo2,thermo3}. Under this framework, many issues such as information paradox arises and discussed \cite{paradox}.
	
	Very recently, based on Duan's topological current $\phi$-mapping theory \cite{Duan}, Wei et. al. \cite{Liu22,Liu21} innovatively proposed that black hole can be viewed as topological thermodynamic defects, and successfully classify different black hole solutions with their global topological charges. The black hole solutions can be divided into three different topological classes according to their different topological numbers. This method offers us an new perspective into black hole solutions and their thermodynamic property.
	
	Along this line, many works continues to discuss the topological classes of different types of black hole. To go beyond the static solutions, in \cite{Kerr}, stationary black holes are considered. The topological number of Kerr and Kerr-Newman black hole are calculated respectively. Moreover, they also calculate the singly-rotating black hole in higher dimensions as well as in Anti-de-Sitter case \cite{Kerr,Kerrads}. In addition to research on topology classes in general relativity, some researchers are also extend these classifications into modified gravity theories such as Gauss-Bonnet gravity \cite{GB} or Lovelock gravity \cite{Lovelock}. The topological number are quite different in these cases, which in turn provide us a new perspective to consider the difference between GR and modified gravity.
	
	Since the study of the topological classifications of black holes
are still in its infancy and the topological number of black holes de-Sitter$($dS$)$ spacetime remains virgin territory, it deserves to be explored deeply. On the other hand, our current universe is in a state of accelerated expansion. While the simplest explanation of such accelerated expansion is a positive cosmological constant \cite{cosmologicalconstant}. More importantly, in dS spacetime there are two kinds of horizon, both emitting Hawking radiation \cite{Teitelboim1,Teitelboim2}, as is different from the thermodynamics of flat or anti de-Sitter$($AdS$)$ spacetime. Thus investigating dS spacetime has theoretical and practical significance \cite{KerrCFT,Zhang19a}. Extending these approach to dS case may provide us some new insight into these universal phenomenon. With these strong motivations in hand, in this paper, we follow this newly-hewed path to study the topological classes of black holes in dS spacetime.

   At the same time, many works about modified gravity, such as adding higher curvature term in GR or considering higher dimension, have been done to reconcile GR with other quantum theories. As a result, de-Sitter spacetime also existed in many modified gravity theories. In this article we will concentrate on thermodynamics of dS black hole including its event horizon and cosmological horizon which can be seen as two different thermodynamic systems \cite{Teitelboim2} and then calculate the topological charge when the angular momentum and electric charge are taken different values. To get more information, we subsequently explore dS black hole in different background, e.g. Gauss-Bonnet (GB) gravity and high dimensional case.
	
	This paper is organized as follows: In section 2, we briefly review some useful results of thermodynamics of asymptotically de-Sitter spacetimes. Then we use the generalized free energy to establish a parameter space, and find the zero points with their topological charge. In section 3, we follow the same step and get the topological number of high dimensional dS black hole with or without Gauss-Bonnet term. Finally we summarize our results in section 4 and made a comparison with anti de-Sitter and flat cases.

	\section{TOPOLOGICAL CLASSES OF four dimensional dS BLACK HOLE SOLUTIONS}
	
	\subsection{Thermodynamics of four dimensional dS black hole}
		In this section, we first give a brief review of thermodynamics of dS black hole. We use the generalized off-shell free energy \cite{York} so that we could classify different black hole solutions. This means black holes with the same energy $($and electric charge or angular momentum, if any$)$ can be in different temperature. The Kerr-Newman de-Sitter metric can be written in the Boyer-Lindquist type coordinates as follows \cite{KNdS,Zhang19a}
		\begin{equation}
			d s^2=-\frac{\Delta_r}{R^2}\left(d t-\frac{a}{\Xi} \sin ^2 \theta d \phi\right)^2+R^2\left(\frac{d r^2}{\Delta_r}+\frac{d \theta^2}{\Delta_\theta}\right) \\
			+\frac{\Delta_\theta \sin ^2 \theta}{R^2}\left(a d t-\frac{r^2+a^2}{\Xi} d \phi\right)^2,
		\end{equation}
		where
		\begin{eqnarray}
			&& R^2=r^2+a^2 \cos ^2 \theta, \quad \Xi=1+\frac{a^2}{L^2}, \\
			&& \Delta_r=\left(r^2+a^2\right)\left(1-\frac{r^2}{L^2}\right)-2 m r+q^2, \\
			&& \Delta_\theta=1+\frac{a^2}{L^2} \cos ^2 \theta, \quad \frac{1}{L^2}=\frac{\Lambda}{3} .
		\end{eqnarray}
		Here $m$, $a$ and $q$ respectively represent the mass, rotating parameter and electric charge of dS black hole.  $L$ is the radius of curvature of spacetime and satisfy $L^{2}=3/\Lambda$ with $\Lambda$ being the positive cosmological constant. Moreover, we have to emphasize that $\Lambda$ is no more a fixed value so that the modified Bekenstein-Smarr mass formula still holds in dS black hole thermodynamics \cite{KNdS}. In this paper, we consider $\Lambda$ as a dynamical variable of the system. The  electromagnetic potential reads
		\begin{equation}
		A_{t}=\frac{q r}{R^{2}}, \quad A_{\phi}=\frac{q r}{R^{2} \Xi} a \sin ^{2} \theta .
	\end{equation}
		The metric is singular where $\Delta_{r}$ vanishes. The algebraic equation $\Delta_{r}=0$ has the four roots which are three positive and one negative solutions in the condition that the relation
		\begin{equation}
			{\left[\left(L^{2}-a^{2}\right)^{2}-12 L^{2}\left(a^{2}+q^{2}\right)\right]^{3}>}{\left[\left(L^{2}-a^{2}\right)^{3}+36 L^{2}\left(L^{2}-a^{2}\right)\left(a^{2}+q^{2}\right)-54 m^{2} L^{4}\right]^{2}}\label{condition}
		\end{equation}
		 The largest positive solution is the cosmological horizon $r_{c}$, the smallest positive solution is inner black hole horizon, and the other positive solution is the black hole event horizon $r_{h}$. The negative solution has no physical meaning. In this paper, we assume that Eq.\eqref{condition} is satisfied, and only focus on the event horizon and cosmological horizon. Interestingly, when using the Euclidean solution to investigate thermodynamics, one finds that the imaginary time periods required to avoid conical singularity at the two horizons do not match at all. It means that the two horizons are in different thermal state and we should discuss them separately. In fact, when one observes either one of two horizons as thermodynamical system, then the other should be viewed as a boundary \cite{Teitelboim2} where the parameters are fixed and there will be no field equations to satisfy. Before the discussion in detail, we first recall the relation between black hole parameters and their corresponding thermodynamical quantities \cite{KNdS}
		 \begin{eqnarray}
		 		M_{h}=\frac{m}{\Xi^{2}}, \quad J_{h}=\frac{m a}{\Xi^{2}}, &&\quad Q_{h}=\frac{q}{\Xi}, \quad \Lambda_{h}=\Lambda,\\
		 	M_{c}=-\frac{m}{\Xi^{2}}, \quad J_{c}=-\frac{m a}{\Xi^{2}}, &&\quad Q_{c}=-\frac{q}{\Xi}, \quad \Lambda_{c}=-\Lambda,
		 \end{eqnarray}
	where subscript $h$ means that it is the physical quantity associated with the black hole event horizon and $c$ for cosmological horizon. The entropy of the two thermodynamic systems reads respectively
	\begin{eqnarray}
		S_{h}=\frac{\pi(r_{h}^{2}+a^2)}{\Xi}, \quad S_{c}=\frac{\pi(r_{c}^{2}+a^2)}{\Xi}.
		\end{eqnarray}
	Through it we could obtain the	generalized Helmholtz free energy as follow \cite{York}
	\begin{eqnarray}
			F_{h}&=&M_{h}-\frac{S_{h}}{\tau_{h}}=\frac{a^2 L^2-a^2 r_{h}^2+L^2 q^2+L^2 r{h}^2-r_{h}^4}{2 L^2 r_{h} \left(\frac{a^2}{L^2}+1\right)^2}-\frac{\pi  \left(a^2+r_{h}^2\right)}{\tau_{h}  \left(\frac{a^2}{L^2}+1\right)},\label{Fh}\\
				F_{c}&=&M_{c}-\frac{S_{c}}{\tau_{c}}=-\frac{a^2 L^2-a^2 r_{c}^2+L^2 q^2+L^2 r{c}^2-r_{c}^4}{2 L^2 r_{c} \left(\frac{a^2}{L^2}+1\right)^2}-\frac{\pi  \left(a^2+r_{c}^2\right)}{\tau_{c}  \left(\frac{a^2}{L^2}+1\right)}.
		\label{Fc}
	\end{eqnarray}
	It is off-shell except at $\tau=1/T$ which means the black hole is in the maximal mixed state.

	\subsection{Topological classes of dS black hole event horizon}
		When the generalized free energy is in hand, we could apply the method in \cite{Liu22} to estabilish a parameter space and find the zero point of the vector field in it. Profoundly, the zero points are exactly corresponding to the on-shell black hole solution. We can calculate the topological number of them by virtue of Duan'$\phi$-mapping topological current theory \cite{Duan}. The number can be seen as a characteristic value of the on-shell black hole solution. Following the spirit of \cite{Liu22}, we define the vector as
	\begin{equation}
		\phi=(\phi^{r_{h}},\phi^{\theta})=(\frac{\partial F}{\partial r_{h}},-\cot\theta \csc\theta)
		\label{vectorfield}
	\end{equation}
	with $\theta\in[0,\pi]$ for convenience. The vector field is on $\theta-r_{h}$ space, and we can see that $\phi^{\theta}$ is divergent when $\theta=0,\pi$, making the direction of vectors point vertically outward at this boundary. The zero point, corresponding to $\tau=1/T$ \cite{Liu21}, can only be obtained when $\theta=\pi/2$.
	Now we introduce the topological current as
	\begin{equation}
		j^\mu=\frac{1}{2 \pi} \epsilon^{\mu \nu \rho} \epsilon_{a b} \partial_\nu n^a \partial_\rho n^b, \quad \mu, \nu, \rho=0,1,2
	\end{equation}
	where $\partial_\nu=\left(\partial / \partial x^\nu\right)$ and $x^\nu=\left(\tau, r_h, \theta\right)$. The unit vector is defined as $n^a=\left(\phi^a /\|\phi\|\right)(a=1,2)$. The conservation law of the current, $\partial_\mu j^\mu=0$ is obvious according to the definition of $j^\mu$, and $\tau$ here serves as a time parameter of the topological defect. By using the Jacobi tensor $\epsilon^{a b} J^\mu(\phi / x)=$ $\epsilon^{\mu \nu \rho} \partial_\nu \phi^a \partial_\rho \phi^b$ and the two-dimensional Laplacian Green function $\Delta_{\phi^a} \ln \|\phi\|=2 \pi \delta^2(\phi)$, the topological current can be written as
	\begin{equation}
		j^\mu=\delta^2(\phi) J^\mu\left(\frac{\phi}{x}\right) .
	\end{equation}
	where $j^\mu$ is nonzero only at $\phi^a\left(x^i\right)=0$, and we denote its $i$-th solution as $\vec{x}=\vec{z}_i$. The topological current density then reads \cite{density}
	\begin{equation}
		j^0=\sum_{i=1}^N \beta_i \eta_i \delta^2\left(\vec{x}-\vec{z}_i\right),
	\end{equation}
	where $\beta_i$ is Hopf index, which counts the number of the loops that $\phi^a$ makes in the vector $\phi$ space when $x^\mu$ goes around the zero point $z_i$. Thus Hopf index is always positive. $\eta_i$ is the Brouwer degree and satisfy $\eta_i=$ $\operatorname{sign}\left(J^0(\phi / x)_{z_i}\right)=\pm 1$. Given a parameter region $\Sigma$, the corresponding topological number can be obtained as
	\begin{equation}
		W=\int_{\Sigma} j^0 d^2 x=\sum_{i=1}^N \beta_i \eta_i=\sum_{i=1}^N w_i,
		\label{W}
	\end{equation}
	where $w_i$ is the winding number for the $i$-th zero point of $\phi$ contained in $\Sigma$ and its value does not depend on the shape of the region where we perform the calculation. Usually, distinct zero points of the vector field are isolated, making Jacobian $J_0(\phi / x) \neq 0$. If Jacobian $J_0(\phi / x) = 0$, it means that the defect bifurcates \cite{Duan2}. Eq. \eqref{W} shows that in any given region, the global topological number is the sum of the winding number of each zero point which reflects the local property of the topological defect.	Based on the approach introduced above, we would investigate dS black hole with different horizons.
	
\begin{figure}
		\centering
		\includegraphics{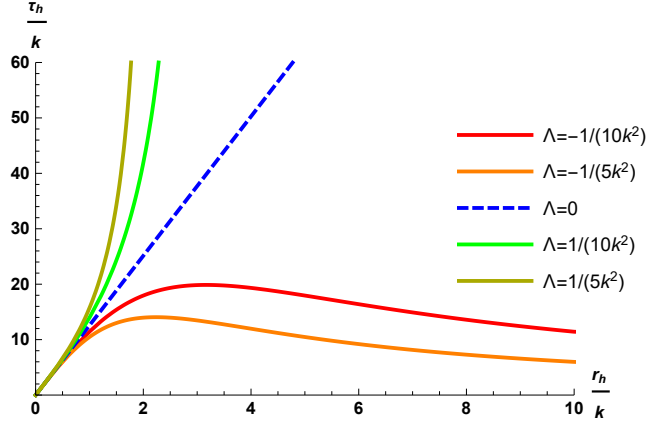}
		\caption{Solution curves in $\tau_{h}-r_{h}$ plane with different values of $\Lambda$, where $k$ is an arbitrary positive constant. It can be seen that for dS black hole the curves are always monotonically increasing which is the same as $\Lambda=0$ case. But for AdS black hole, the curve go down to zero so there are two zero points for a given $\tau$
		}
		\label{h1-1}
	\end{figure}

\begin{figure}
		\centering
		\includegraphics{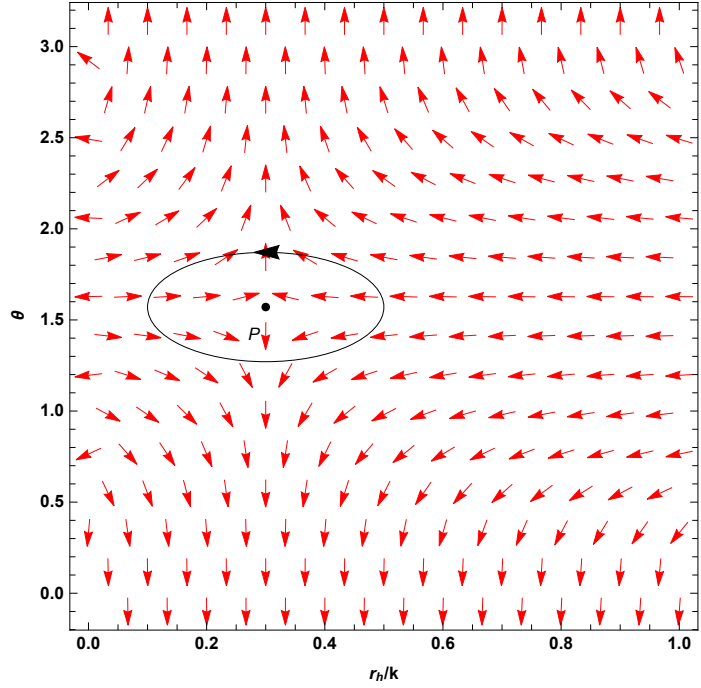}
		\caption{
			Unit vector field in parameter space with  $a=0$, $q=0$ and $\Lambda=3/k^{2}$ Schwarzchild-dS solution as $\tau_{h}\approx5.16 k$. $P$ marked with black dot at $(r_{h}/k,\theta)=(0.3,\pi/2)$ is the zero point of the vector field.  The black contour is a closed loop enclosing the zero point and we can performing the calculation of Eq.\eqref{W} inside it. The shape and the size of the loop will not affect the winding number.
		}
		\label{h1-2}
	\end{figure}

Next we consider dS black hole as a thermodynamical object in a "box", the boundary of which is at the cosmological horizon $r_{c}$. We first concentrate on $a=0$ and $q=0$ case whose generalized free energy is reduced to
	\begin{equation}
		F_{h}=M_{h}-\frac{S_{h}}{\tau_{h}}=\frac{L^2 r_{h}-r_{h}^3}{2 L^2 }-\frac{\pi  r_{h}^2}{\tau_{h} }.
	\end{equation}
	We define the vector field as
	\begin{equation}
		\phi=(\phi^{r_{h}},\phi^{\theta})=(-\frac{3 r_h^2}{2 L^2}-\frac{2 \pi  r_h}{\tau_{h} }+\frac{1}{2},-\cot\theta \csc\theta).
	\end{equation}
	By solving the equation $\phi=0$, we acquire the relation
	\begin{equation}
		\tau_{h}=\frac{4 \pi  L^2 r_h}{L^2-3 r_h^2}, \quad \theta=\frac{\pi}{2}.
	\end{equation}
	We take different values of $\Lambda$ and depict the on-shell solution curve on $\tau_{h}-r_{h}$ plane as shown in Fig. \ref{h1-1}. When $\tau_{h}\approx5.16 k$, different from Schwarzchild-AdS case, there would be one zero point in $\theta-r_{h}$ plane at $(r_{h}/k,\theta)=(0.3,\pi/2)$, which is illustrated in Fig. \ref{h1-2}.  The loop surrounding the zero point sets the boundary of a given region so we can use Eq. \eqref{W} to acquire the winding number of point $P$. In this case, it can be easily calculated as $w=-1$. Since there is only one defect in the parameter space, it gives rise the global topological number $W=-1$ for non-rotating and uncharged Schwarzchild-dS black hole solution.

	Now we would further explore the rotating cases ($a\neq 0, q=0$). The generalized free energy reads
	\begin{equation}
		F_{h}=\frac{-a^2 r_h^2+a^2 L^2+L^2 r_h^2-r_h^4}{2 L^2 \left(\frac{a^2}{L^2}+1\right)^2 r_h}-\frac{\pi  \left(a^2+r_h^2\right)}{\tau_{h}  \left(\frac{a^2}{L^2}+1\right)}.
	\end{equation}
	As a result, we define the vector field as
	\begin{equation}
		\phi=(\phi^{r_{h}},\phi^{\theta})=(-\frac{L^2 \left(\tau_{h}  \left(a^2-L^2\right) r_h^2+4 \pi  \left(a^2+L^2\right) r_h^3+a^2 L^2 \tau_{h} +3 \tau_{h}  r_h^4\right)}{2 \tau_{h}  \left(a^2+L^2\right)^2 r_h^2},-\cot\theta \csc\theta).
	\end{equation}
	Solving the equation $\phi=0$ give us
	\begin{equation}
		\tau_{h}=-\frac{4 \pi  \left(a^2+L^2\right) r_h^3}{\left(a^2-L^2\right) r_h^2+a^2 L^2+3 r_h^4}, \quad \theta=\frac{\pi}{2}.
	\end{equation}
	
\begin{figure}
		\centering
		\includegraphics{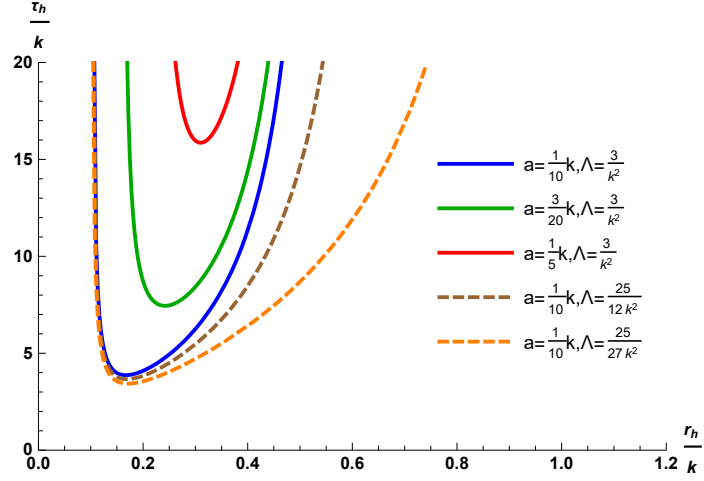}
		\caption{Solution curves in $\tau-r_{h}$ plane with different values of $\Lambda$ and $a$. There is always one turning point in the curve as long as $a\neq0$.
		}
		\label{h2-1}
	\end{figure}

\begin{figure}
		\centering
		\includegraphics{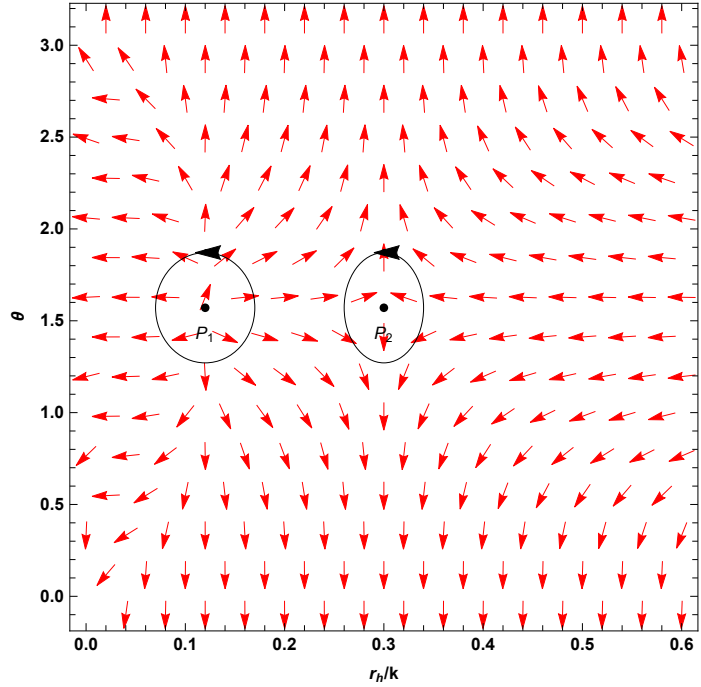}
		\caption{Vector field in $\theta-r_{h}$ plane with $a\neq0$. We take $\tau=6.25k$. The points marked in black dot are zero points of the field. They are respectively at $P_{1}=(r_{h}/k,\theta)=(0.12,\pi/2)$ and $P_{2}=(r_{h}/k,\theta)=(0.3,\pi/2)$.
		}
		\label{h2-2}
	\end{figure}

	Upon the value of $\Lambda$ and $a$ are determined, we could acquire a curve in $\tau_{h}-r_{h}$ plane. For instance, as shown in Fig. \ref{h2-1}, when taking $\Lambda=3/k^{2}$ and $a=k/10$, the blue curve goes down rapidly to a peak at $(r_{h},\tau_{h})\approx(0.17k,3.87k)$ and climbs as $r_{h}$ gets larger. Consequently, the vector field possesses two zero points for large $\tau_{h}$, in contrast to one zero point for Schwarzschild-dS black hole case. As $\tau_{h}=6.25k$, the two intersection points are respectively at $r_{h}\approx0.12k$ and $r_{h}\approx0.3k$. We illustrate the vector field and the zero points in Fig. \ref{h2-2}. When $\tau=\tau_{cri}\approx3.87k$, the intersection points coincide, and for smaller $\tau_{h}$, annihilate. It is easy to check the critical point satisfy $d^{2}\tau_{h}/dr_{h}^{2}>0$, which belongs to generation point. For $\tau>\tau_{cri}$, we find that the winding number of the two zero points are $w_{1}=1$ and $w_{2}=-1$. Thus the global topological number for Kerr-dS solution is $W=w_{1}+w_{2}=0$, different from non-rotating case.

\begin{figure}
		\centering
		\includegraphics{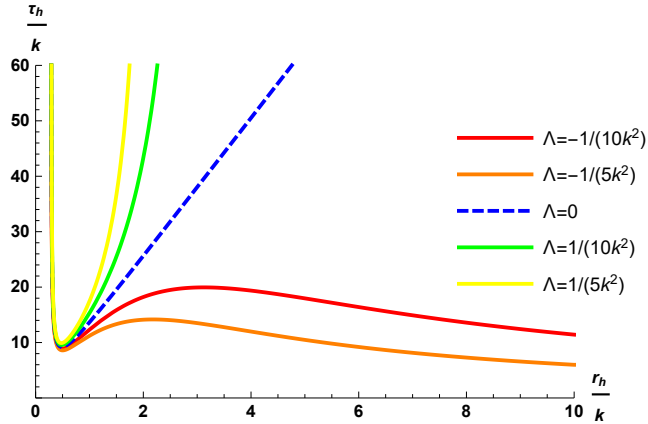}
		\caption{On-shell charged solution curves for different cosmological constant. The trend of the curves are the same when $r_{h}$ is small, but look different as $r_{h}$ gets larger.
		}
		\label{d=4RN}
	\end{figure}

	When taking electric charge into account, the generalized off-shell free energy of dS black hole is given by Eq. \eqref{Fh}. Following the same step, we get the on-shell solution curve in $\tau_{h}-r_{h}$ plane, which satisfy
	\begin{equation}
		\tau_{h}=-\frac{4 \pi  \left(a^2+L^2\right) r_h^3}{\left(a^2-L^2\right) r_h^2+L^2 \left(a^2+q^2\right)+3 r_h^4}.
	\end{equation}
	\begin{figure}
	\centering
	\includegraphics{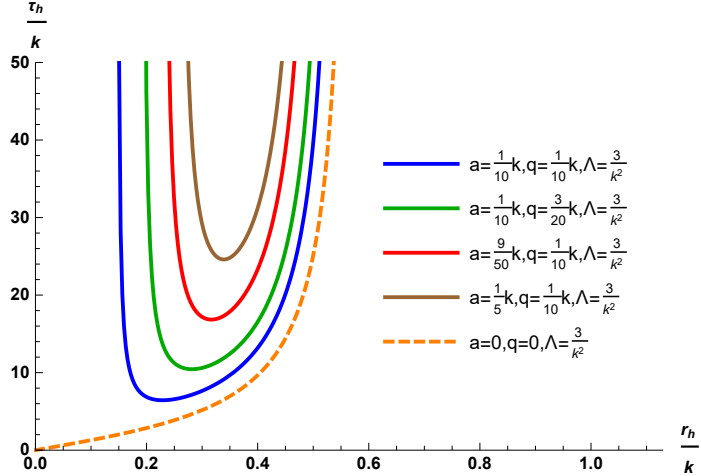}
	\caption{Solution curves in $\tau_{h}-r_{h}$ plane with different values of $\Lambda$, $a$ and $q$. We found that non-trivial $q$ do not bring essential difference to the trend of the curve compared with $a\neq0$ but $q=0$ case.
	}
	\label{h4-1}
\end{figure}
\begin{figure}
	\centering
	\includegraphics{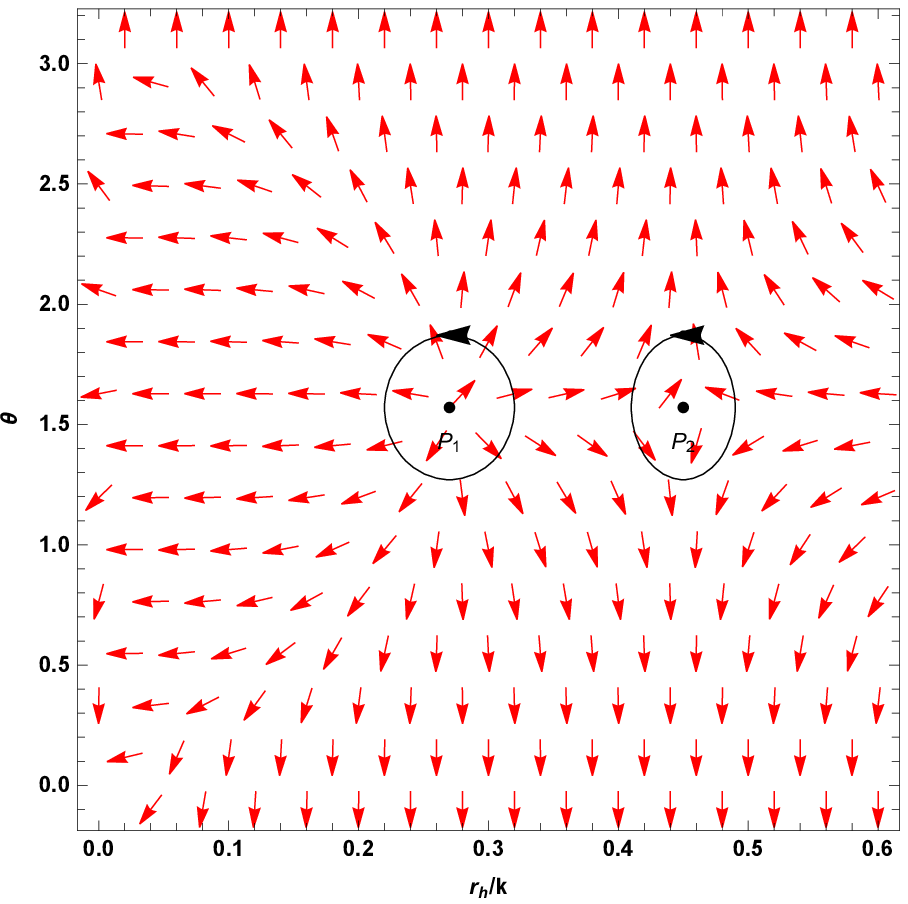}
	\caption{Vector field in $\theta-r_{h}$ plane with $a\neq0$ and $q\neq0$. We take $\tau_{h}\approx55.8k$. The points marked in black dot are zero points of the field. They are respectively at $P_{1}=(r_{h}/k,\theta)\approx(0.27,\pi/2)$ and $P_{2}=(r_{h}/k,\theta)\approx(0.45,\pi/2)$.
	}
	\label{h4-2}
\end{figure}
	Interestingly, we find that the electric charge do not change the rough trend of the curve. As shown in Fig. \ref{h4-1}, there is invaribly one generation point in $\tau_{h}-r_{h}$ plane, and $\tau_{cri}$ merely gets larger as the value of $q$ taken larger. There are two zero points as well. We calculate the winding number of them for a given $\tau_{h}$ and find $w_{1}=1$, $w_{2}=-1$. See Fig. \ref{h4-2} as an example. The global topological number of Kerr-Newman-dS black hole solution is $W=0$, which is the same as the Kerr-dS black hole. We also calculate the non-rotating charged cases whose global topological number is $W=w_{1}+w_{2}=0$. Hence from the perspective of topological charge, these three kinds of black hole are just the same. Moreover, compared with negative cosmological constant case, for instance, the charged Reissner-Nordstrom anti de-Sitter (RN-AdS) black hole in four dimension \cite{Liu22}, RN-dS black hole solution evidently has fewer zero points and the topological number are distinct. We plot their on-shell solution curves in Fig. \ref{d=4RN}. We could conclude the sign of cosmological constant may have an influence on the topological number and quantity of defects of black hole of the same type.
	
	\subsection{Topological classes of dS solutions with cosmological horizon}
	In the following we shall discuss the thermodynamics and the corresponding topological charge of cosmological horizon. Now the black hole event horizon $r_{c}$ serve as a boundary, which plays the analogous role of coordinate origin of empty de Sitter space \cite{KNdS}. The form of generalized free energy of cosmological horizon resembles the black hole event horizon counterpart. For $a=0$, $q=0$ case, it reads
		\begin{equation}
		F_{c}=-\frac{L^2 r_c^2-r_c^4}{2 L^2 r_c}-\frac{\pi  r_c^2}{\tau _c}.
	\end{equation}
	The vector field can be defined as
	\begin{equation}
		\phi=(\phi^{r_{c}},\phi^{\theta})=(\frac{3 r_c^2}{2 L^2}-\frac{2 \pi  r_c}{\tau _c}-\frac{1}{2},-\cot\theta \csc\theta).
	\end{equation}
	By solving the equation $\phi=0$, we acquire the relation
	\begin{equation}
		\tau_{c}=-\frac{4 \pi  L^2 r_c}{L^2-3 r_c^2}.
		\label{tauc}
	\end{equation}
\begin{figure}
	\centering
	\includegraphics{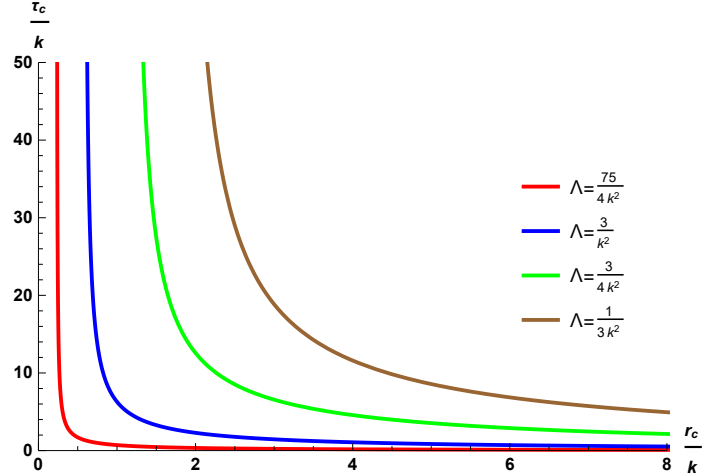}
	\caption{Solution curves in $\tau_{c}-r_{c}$ plane with different values of $\Lambda$. It can be seen that the curves are always monotonically decreasing.
	}
	\label{c1-1}
\end{figure}
\begin{figure}
	\centering
	\includegraphics{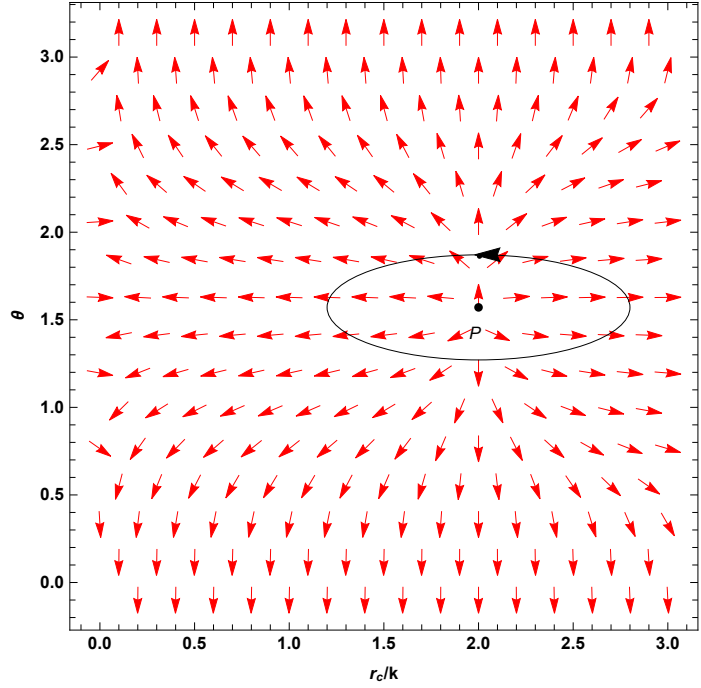}
	\caption{
		Unit vector field in parameter space with  $a=0$, $q=0$ and $\Lambda=3/k^{2}$ solution as $\tau_{c}\approx8\pi k/11$. $P$ marked with black dot at $(r_{c}/k,\theta)=(2,\pi/2)$ is the zero point of the vector field.  We performing the calculation of Eq.\eqref{W} inside the black contour and obtain the winding number $w=1$.
	}
	\label{c1-2}
\end{figure}
We could see that the only difference of the $r-\tau$ relation with black hole event horizon is the sign of Eq. \eqref{tauc}. But it makes the winding number distinct as we will show. We take various value of $\Lambda$ and depict the on-shell solution curve on $\tau_{c}-r_{c}$ plane as shown in Fig. \ref{c1-1}. When $\tau_{c}\approx8\pi k/11$, there would be one zero point in $\theta-r_{c}$ plane at $(r_{c}/k,\theta)=(2,\pi/2)$, which is illustrated in Fig. \ref{c1-2}.  By virtue of Eq.\eqref{W}, we get the winding number of the single zero point $w=1$ which is different from non-rotating uncharged black hole event horizon case.

\begin{figure}
	\centering
	\includegraphics{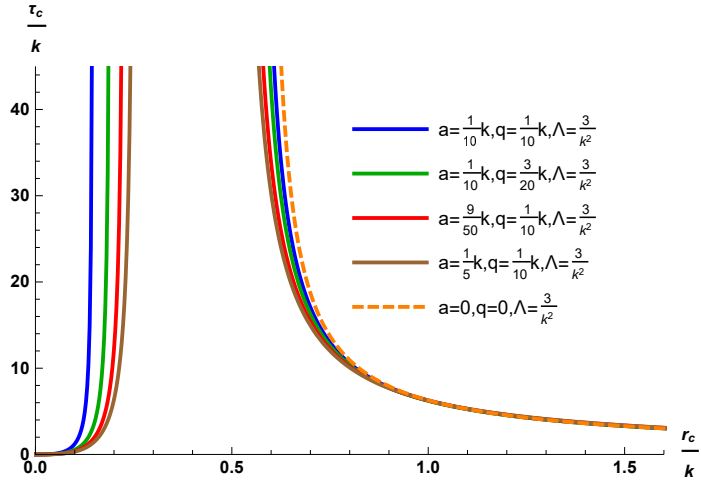}
	\caption{Solution curves in $\tau_{c}-r_{c}$ plane with different values of $\Lambda$.
	}
	\label{c4-1}
\end{figure}
\begin{figure}
	\centering
	\includegraphics{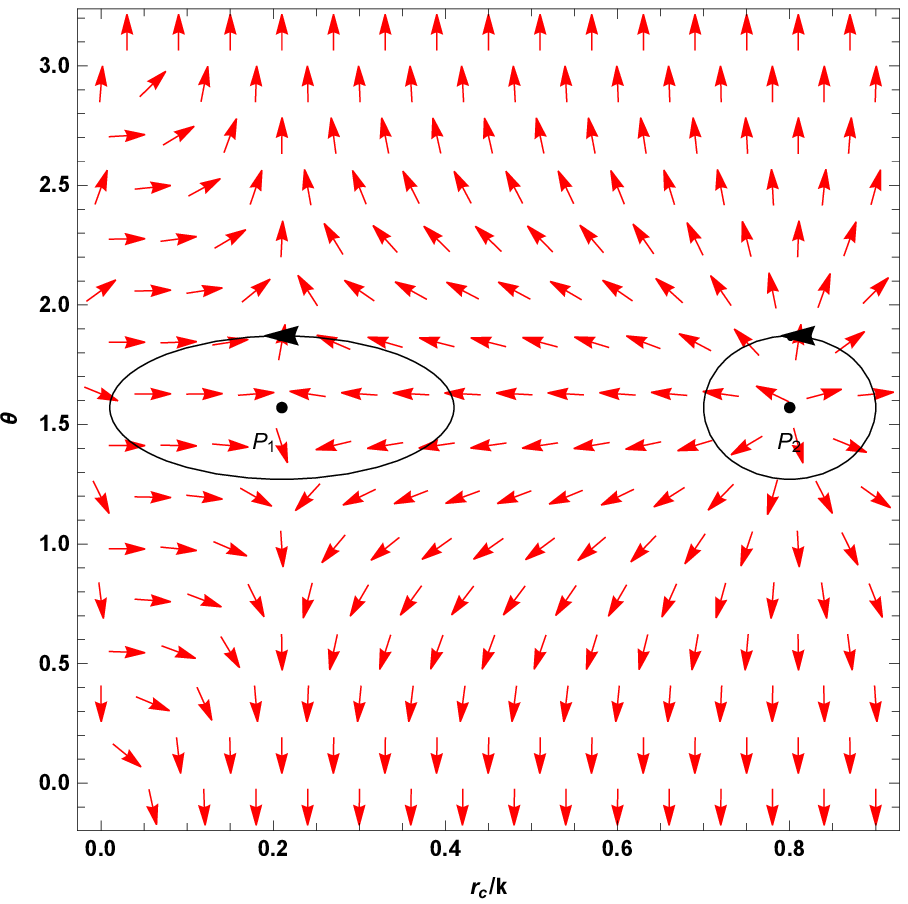}
	\caption{
		Unit vector field in parameter space with  $a=k/5$, $q=k/10$ and $\Lambda=3/k^{2}$ solution as $\tau_{c}\approx10k$. $P_{1}$ and $P_{2}$ are marked with black dot at $(r_{c}/k,\theta)\approx(0.21,\pi/2)$ and $(r_{c}/k,\theta)\approx(0.81,\pi/2)$. We find that $w_{1}=-1$ and $w_{2}=1$. It makes the global topological number $W=0$.
	}
	\label{c4-2}
\end{figure}

For $a\neq0$, $q\neq0$ or rotating and charged case, we find that there are invariably two zero points in vector field.  Although the global topological charge is $W=0$ again, the winding number of defects with smaller radius, $w_{1}=-1$, or with larger radius, $w_{2}=1$ is respectively different with black hole event horizon case. We plot the solution curve in Fig. \ref{c4-1} and calculate the topological charge when taking $\tau_{c}=10k$, $a=k/5$,$q=k/10$ as well as $\Lambda=3/k^{2}$ as an example, as shown in Fig. \ref{c4-2}.
	
	
		\section{TOPOLOGICAL CLASSES OF higher dimensional dS BLACK HOLE SOLUTIONS with Gauss-Bonnet term}
	
\begin{figure}
	\centering
	\includegraphics{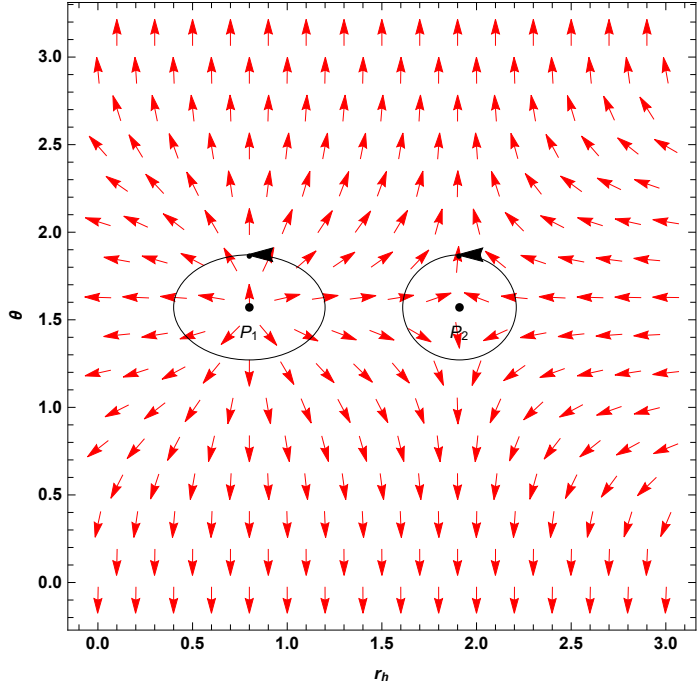}
	\caption{Vector field of five dimensional uncharged dS black hole when taking $\tau_{h}\approx21.18$ and $\Lambda=1/10$. The two zero points are at $(r_{h},\theta)\approx(0.8,\pi/2)$ and $(r_{h},\theta)\approx(1.91,\pi/2)$, marked in black dots. The winding number are $w_{1}=1$ and $w_{2}=-1$.
	}
	\label{GBSch}
\end{figure}

	To gain more knowledge of topological classes of dS black hole, we further explore high dimensional solutions and add higher derivative curvature terms in the form of Gauss-Bonnet gravity. For simplicity we only focus on static black hole horizon. The most general $d$-dimensional spherically symmetrical metric solution reads \cite{Cai,GBdS}
	\begin{equation}
		d s^{2}=-f(r) d t^{2}+\frac{d r^{2}}{f(r)}+r^{2} d \Omega_{d-2}^{2},
	\end{equation}
	where
	\begin{eqnarray}
		f(r)&=&1+\frac{r^{2}}{2 \alpha}-\frac{r^{2-d / 2} \sqrt{r^{d}+4 \alpha \left(r \omega_{d-3}-q^{2} r^{4-d}-r^{d} / L^{2}\right)}}{2 \alpha},\\
		\omega_{d-3}&=&\frac{16 \pi G m}{(d-2) \Omega_{d-2}}
	\end{eqnarray}
	$\alpha$ represents the Gauss-Bonnet coefficient and $L^2=(d-1) (d-2)/2 \Lambda$. The black hole horizon $r_{h}$ can also be obtained when $f(r)=0$. By virtue of gravitational path integral we could acquire the generalized free energy of charged Gauss-Bonnet black hole in arbitrary dimensions as
		\ba
		F_{h}&=&\frac{\Omega _{d-2} r_h^{-d-5} \left(2 \pi  (d-3) r_h^{2 d} \left(\left(d^2-3 d+2\right) \left(\alpha +r_h^2\right)-2 \Lambda  r_h^4\right)+(d-1) q^2 r_h^8\right)}{16 \pi ^2 \left(d^2-4 d+3\right)}\nn\\
&&-\frac{\Omega _{d-2} r_h^{d-2} \left(\frac{2 \alpha  (d-2)}{(d-4) r_h^2}+1\right)}{2 \tau _h}.
		\label{FGB}
	\ea
	In what follows we would use it to explore the topological classes of high dimensional dS static black hole.	
	We first investigate uncharged Gauss-Bonnet dS cases. Under this circumstance, the generalized free energy Eq. \eqref {FGB} is reduced to
	\ba
F_{h}&=&\frac{(d-2) \Omega _{d-2} r_h^{d-3} \left(-\frac{2 \Lambda  r_h^2}{(d-2) (d-1)}+\frac{\alpha }{r_h^2}+1\right)}{8 \pi }\nn\\
&&-\frac{\Omega _{d-2} r_h^{d-2} \left(\frac{2 \alpha  (d-2)}{(d-4) r_h^2}+1\right)}{2 \tau _h}
	\ea
		Following the same step, the vector field can be defined as Eq.\eqnref{vectorfield}. So we have
	\begin{equation}
		\phi^{r_{h}}=-\frac{r_{h}^{d-6} \Omega _{d-2} \left(\tau _h \left(-\alpha  \left(d^2-7 d+10\right)-\left(d^2-5 d+6\right) r_{h}^2+2 \Lambda  r_{h}^4\right)+4 \pi  (d-2) r_{h} \left(2 \alpha +r_{h}^2\right)\right)}{8 \pi  \tau _h}.
	\end{equation}
\begin{figure}
	\centering
	\includegraphics{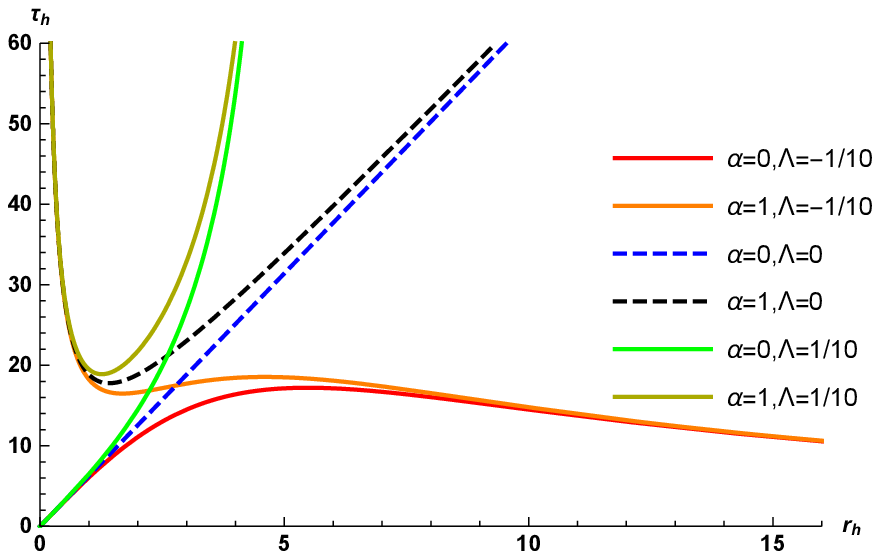}
	\caption{On-shell solution curve of uncharged black hole for different $\alpha$ in AdS or dS spacetime of dimension $d=5$.
	}
	\label{d=5Sch}
\end{figure}
\begin{figure}
	\centering
	\includegraphics{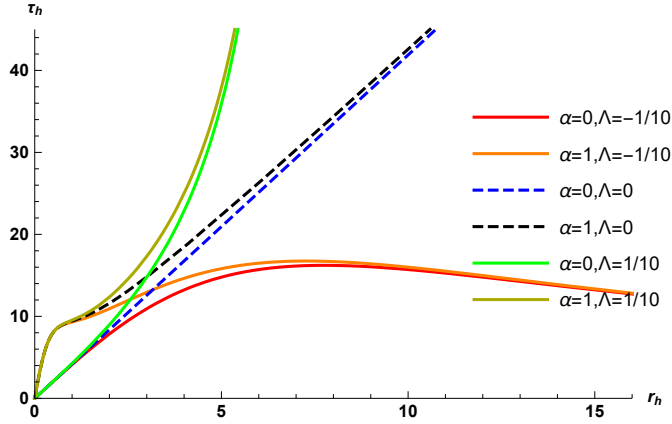}
	\caption{On-shell solution curve of uncharged black hole for different $\alpha$ in AdS or dS spacetime of dimension $d=6$.
	}
	\label{d=6Sch}
\end{figure}
Solving the equation $\phi^{r_{h}}=0$, we get the on-shell solution curves on $\tau_{h}-r_{h}$ plane for the given value of $\alpha$ and dimensions, as illustrated in Figs. \ref{d=5Sch} and \ref{d=6Sch}. We could then calculate the topological charge for the on-shell solution. For instance, when $\alpha=0$ and $d=5$, we find that there is always one zero point whose winding number is $w=-1$. Whereas taking higher curvature term into consideration, the vector field of uncharged dS black hole would possess two zero points, whose winding number are $w_{1}=1$ and $w_{2}=-1$ respectively. Thus the total topological number is 0, in contrast to 1 in AdS case. We depict the vector field of five dimensional dS black hole in Fig. \ref{GBSch} as $\tau_{h}\approx21.18$.

Obviously, for uncharged black hole, the number of zero points change when introducing Gauss-Bonnet term, as is different from the charged case \cite{GB}. In addition, the dimension plays an important role in altering the topological number of Gauss-Bonnet dS black hole. However, the most interesting thing is that, compared with asymptotic flat spacetime, a positive cosmological constant do not bring changes to the quantity of zero points and the global topological number no matter whether higher curvature term is added. It is so different from the conclusion we get in spacetime with negtive cosmological constant. Moreover, the charged-GB-dS black hole, the total topological number can also be calculated by using the free energy \eqref{FGB}, and the result turns out to be $W=0$ in five and six dimensions. This is clearly different with the GB-dS black hole which prossess a spacetime dimension dependent total topological number.


	\section{Conclusion}
	In this paper, we use the generalized free energy of dS black hole solution with different horizons to define a vector field in a parameter space $\theta-r_{h}$ or $\theta-r_{c}$. We find the zero point of the field and obtain the winding number by applying Duan's $\phi$-mapping topological current theory.

	It is discovered that the dS black hole solution with smaller horizon $r_{h}$ has one zero point for $a=q=0$ and two for rotating and charged cases. The global topological charges are -1 and 0 respectively. Whereas for solution with cosmological horizon $r_{c}$, the topological number of non-rotating uncharged case is 1, inverse of event horizon case. But as $a\neq0$ or $q\neq0$, the number obtained is the same, 0. As well, we discuss the black hole horizon in higher dimension case ($d=5$ or $d=6$) with Gauss-Bonnet term and conclude that static dS black hole are of the same category with static asymptotic flat black hole solution. To have an overview of the results for different black hole solution in general relativity or higher dimensional modified gravity,  we list them in Table \ref{Table1}.
	\begin{table}
		\centering
		\begin{tabular}{|c|c|c|c|c|c|}
			\hline
			Black hole solutions &Number of defects & $W$ &Black hole solutions & Number of defects & $W$ \\
			\hline
			Schwarzschild&	1&	-1& 	Schwarzschild-d$S_{5}$ with $r_{h}$&	1&	-1\\
			\hline
			Kerr&	2&	0&RN-d$S_{5}$ with $r_{h}$&	2&	0\\
			\hline
			RN&	2&	0&GB-d$S_{5}$ with $r_{h}$&	2&	0\\
			\hline
			KN&	2&	0&RN-d$S_{5}$-GB with $r_{h}$&	2&	0\\
			\hline
			Schwarzschild-dS with $r_{h}$&	1&	-1&Schwarzschild-d$S_{6}$ with $r_{h}$&	1&	-1\\
			\hline
			Kerr-dS with $r_{h}$&	2&	0&RN-d$S_{6}$ with $r_{h}$&	2&	0\\
			\hline
			RN-dS with $r_{h}$&	2&	0&GB-d$S_{6}$ with $r_{h}$&	1&	-1\\
			\hline
			KN-dS with $r_{h}$&	2&	0&RN-d$S_{6}$-GB with $r_{h}$&	2&	0\\
			\hline
			Schwarzschild-dS with $r_{c}$&	1&	1&RN-AdS&	3&	1\\
			\hline
			Kerr-dS with $r_{c}$&	2&	0& & & \\
			\hline
			RN-dS with $r_{c}$&	2&	0& & & \\
			\hline
				KN-dS with $r_{c}$&	2&	0& & & \\
			\hline
		\end{tabular}
		\caption{The number of zero points and global topological number for different black hole solutions.}
		\label{Table1}
	\end{table}

	The previous works \cite{Liu22,Kerr} indicate that all black hole solutions in the pure Einstein-Maxwell gravity theory should be classified into three different topological classes for four and higher spacetime dimensions. This observation is further enhanced in this paper. Moreover, there are many issues that deserve further investigation. Generalize our results to stationary cases and compare with the existing result will be interesting. Moreover, another interesting object is to investigate the topological number of the black hole solutions in the supergravity with or without positive/negative cosmological constant. We leave these interesting topics for future studies.

	\begin{acknowledgements}
		This work is supported by NSFC with Grants No.12275087 and ``the Fundamental Research Funds for the Central Universities''.
	\end{acknowledgements}	
	\appendix
	\par
	\bibliographystyle{unsrt}

\end{document}